\documentclass[%
 reprint,
superscriptaddress,
 amsmath,amssymb,
 aps,
 prl,
floatfix,
]{revtex4-2}

\usepackage{graphicx}
\usepackage{dcolumn}
\usepackage{bm}

\usepackage{mhchem}
\usepackage{ragged2e}
\usepackage{caption}
\usepackage{subcaption}
\usepackage{siunitx}
\usepackage{float}
\usepackage[dvipsnames]{xcolor}
\usepackage[colorlinks=true,
            linkcolor=Plum,
            citecolor=Plum,
            urlcolor=Plum]{hyperref}

\DeclareSIUnit{\Rydberg}{Ry}
\graphicspath{{./figures/}}		

\usepackage{pdfpages} 
\usepackage{pgffor} 

\makeatletter
\AtBeginDocument{\let\LS@rot\@undefined}
\makeatother

\newif\ifarXiv

\arXivtrue 
\def\supplementfilename{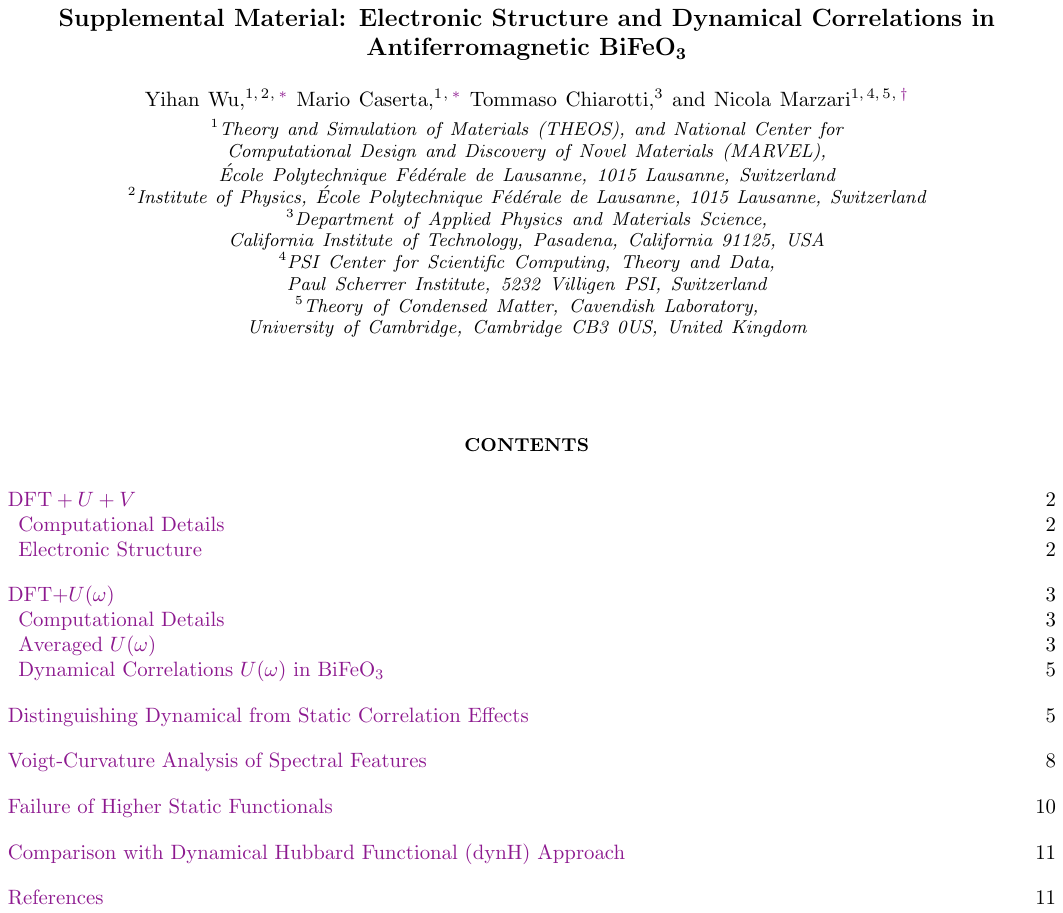}

\pdfximage{\supplementfilename}
\def\numbersupplementpages{\the\pdflastximagepages}

\begin{document}

\preprint{APS/123-QED}

\title{
    Electronic Structure and Dynamical Correlations in Antiferromagnetic BiFeO$_3$
}

\author{Yihan Wu}
\thanks{These authors contributed equally to this work.}
\affiliation{Theory and Simulation of Materials (THEOS), and National Center for Computational Design and Discovery of Novel Materials (MARVEL), École Polytechnique Fédérale de Lausanne, 1015 Lausanne, Switzerland}
\affiliation{Institute of Physics, École Polytechnique Fédérale de Lausanne, 1015 Lausanne, Switzerland}

\author{Mario Caserta}
\thanks{These authors contributed equally to this work.}
\affiliation{Theory and Simulation of Materials (THEOS), and National Center for Computational Design and Discovery of Novel Materials (MARVEL), École Polytechnique Fédérale de Lausanne, 1015 Lausanne, Switzerland}

\author{Tommaso Chiarotti}
\affiliation{Department of Applied Physics and Materials Science, California Institute of Technology, Pasadena, California 91125, USA}

\author{Nicola Marzari}
\email{nicola.marzari@epfl.ch}
\affiliation{Theory and Simulation of Materials (THEOS), and National Center for Computational Design and Discovery of Novel Materials (MARVEL), École Polytechnique Fédérale de Lausanne, 1015 Lausanne, Switzerland}
\affiliation{PSI Center for Scientific Computing, Theory and Data, Paul Scherrer Institute, 5232 Villigen PSI, Switzerland}
\affiliation{Theory of Condensed Matter, Cavendish Laboratory, University of Cambridge, Cambridge CB3 0US, United Kingdom}

\date{\today}

\begin{abstract}
    We study the electronic structure and dynamical correlations in antiferromagnetic BiFeO$_3$, a prototypical room-temperature multiferroic, using a variety of static and dynamical first-principles methods. Conventional static Hubbard corrections (DFT+$U$, DFT+$U$+$V$) incorrectly predict a deep-valence Fe $3d$ peak (around $-7\,\text{eV}$) in antiferromagnetic BiFeO$_3$, in contradiction with hard-X-ray photoemission. We resolve this failure by using a recent generalization of DFT+$U$ to include a frequency-dependent screening – DFT+$U(\omega)$ – or using a dynamical Hubbard functional (dynH). The screened Coulomb interaction $U(\omega)$, computed with spin-polarized RPA and projected onto maximally localized Fe $3d$ Wannier orbitals, is expressed as a sum-over-poles, yielding a self-energy that augments the Kohn–Sham Hamiltonian. This DFT+$U(\omega)$ approach predicts a fundamental band gap of $1.53\,\text{eV}$, consistent with experiments, and completely eliminates the unphysical deep-valence peak. The resulting simulated HAXPES spectrum reproduces the experimental lineshape with an accuracy matching or exceeding that of far more demanding DFT+DMFT calculations. Our work demonstrates the critical nature of dynamical screening in complex oxides and establishes DFT+$U(\omega)$ as a predictive, computationally efficient method for correlated materials.
\end{abstract}

\maketitle


%
\ce{BiFeO3} (bismuth ferrite) is the archetypal room-temperature multiferroic, distinguished by the coexistence of robust ferroelectricity ($T_C \approx \SI{1103}{K}$) and antiferromagnetism ($T_N \approx \SI{643}{K}$)~\cite{teague_dielectric_1970, kiselev_detection_1963, ruette_magnetic-field-induced_2004}. Its distorted $R3c$ perovskite structure~\cite{sosnowska_neutron_2001, kubel_structure_1990}, driven by the stereochemically active Bi $6s^2$ lone pair, hosts a large electric polarization coupled to a complex magnetic cycloid~\cite{das_impedance_2018, ruette_magnetic-field-induced_2004}. This rare simultaneous ordering of charge and spin degrees of freedom establishes \ce{BiFeO3} as a unique platform for fundamental research and a leading candidate for magnetoelectric applications.

Despite extensive experimental characterization, achieving an accurate theoretical description of \ce{BiFeO3}'s electronic structure remains a formidable challenge, primarily due to intricate electronic correlations within the partially filled Fe $3d$ shell~\cite{paul_investigation_2018, neaton_first-principles_2005}. Density-functional theory (DFT) with local and semi-local exchange correlation functionals (i.e., LDA, GGA) notoriously fails, predicting a near-metallic state by severely underestimating the band gap and often misrepresenting orbital occupancies~\cite{neaton_first-principles_2005}. Static mean-field corrections (DFT+$U$, DFT+$U$+$V$) attempt to remedy this by adding a local constant Hubbard-like penalty for on-site and inter-site Coulomb repulsion. A persistent and well-documented issue in all static Hubbard-based treatments (DFT+$U$, DFT+$U$+$V$) of \ce{BiFeO3} is the appearance of an unphysical, sharp peak attributed primarily to Fe $3d$ states at binding energies around \SI{-7}{eV} to \SI{-8}{eV} below the Fermi level~\cite{paul_investigation_2018, inizan_study_2018, mazumdar_valence_2016}. This feature starkly contradicts experimental photoemission spectra, which instead reveals a dip at the corresponding valence band energy~\cite{mazumdar_valence_2016, paul_investigation_2018}. The persistence of this artifact underscores the need to improve static mean-field corrections.
\begin{figure*}[t]
    \centering
    \subfloat[]{%
        \includegraphics[width=0.245\textwidth]{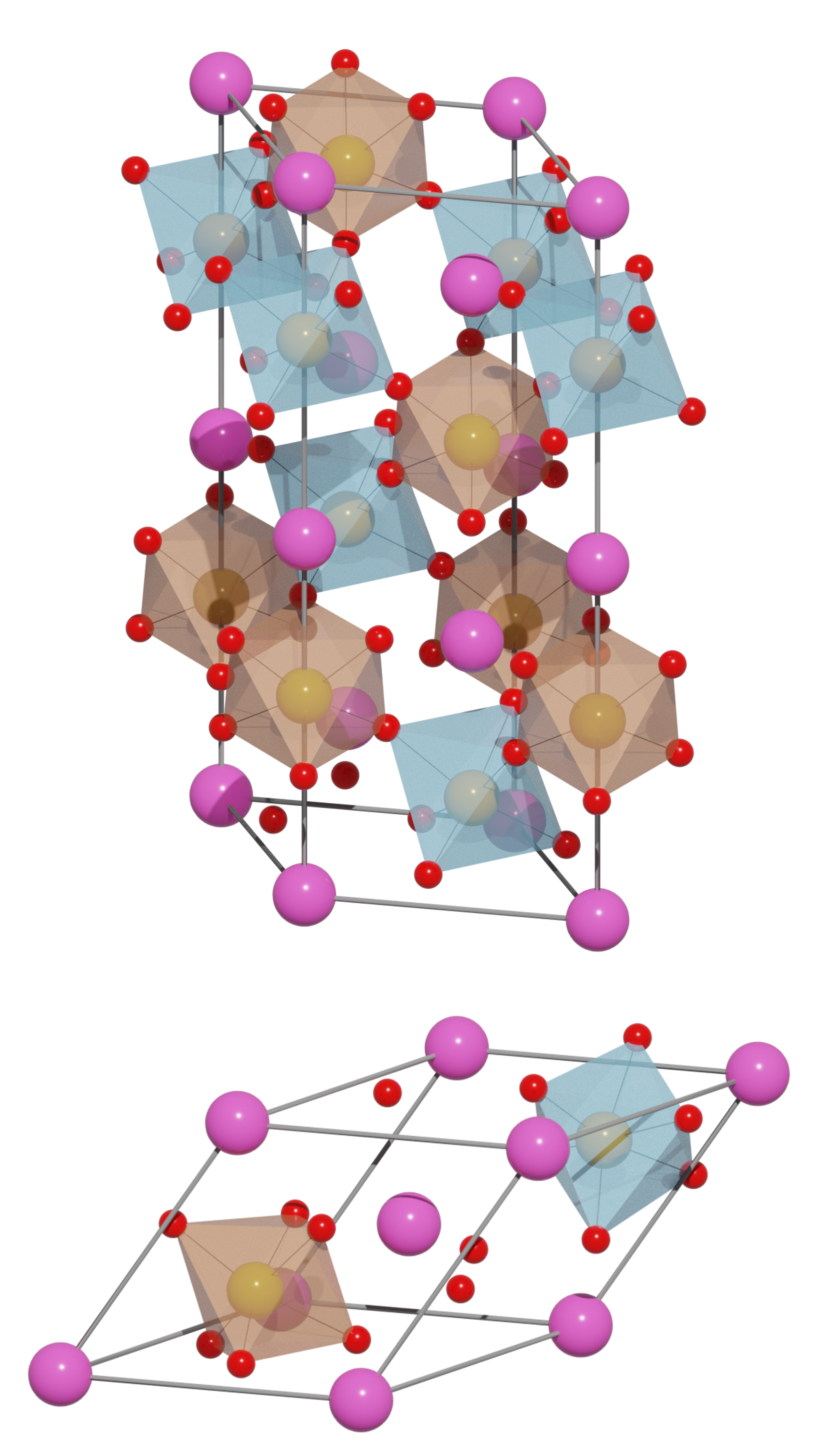}%
        \label{fig:subfigA}%
    }
    \subfloat[]{%
        \includegraphics[width=0.76\textwidth]{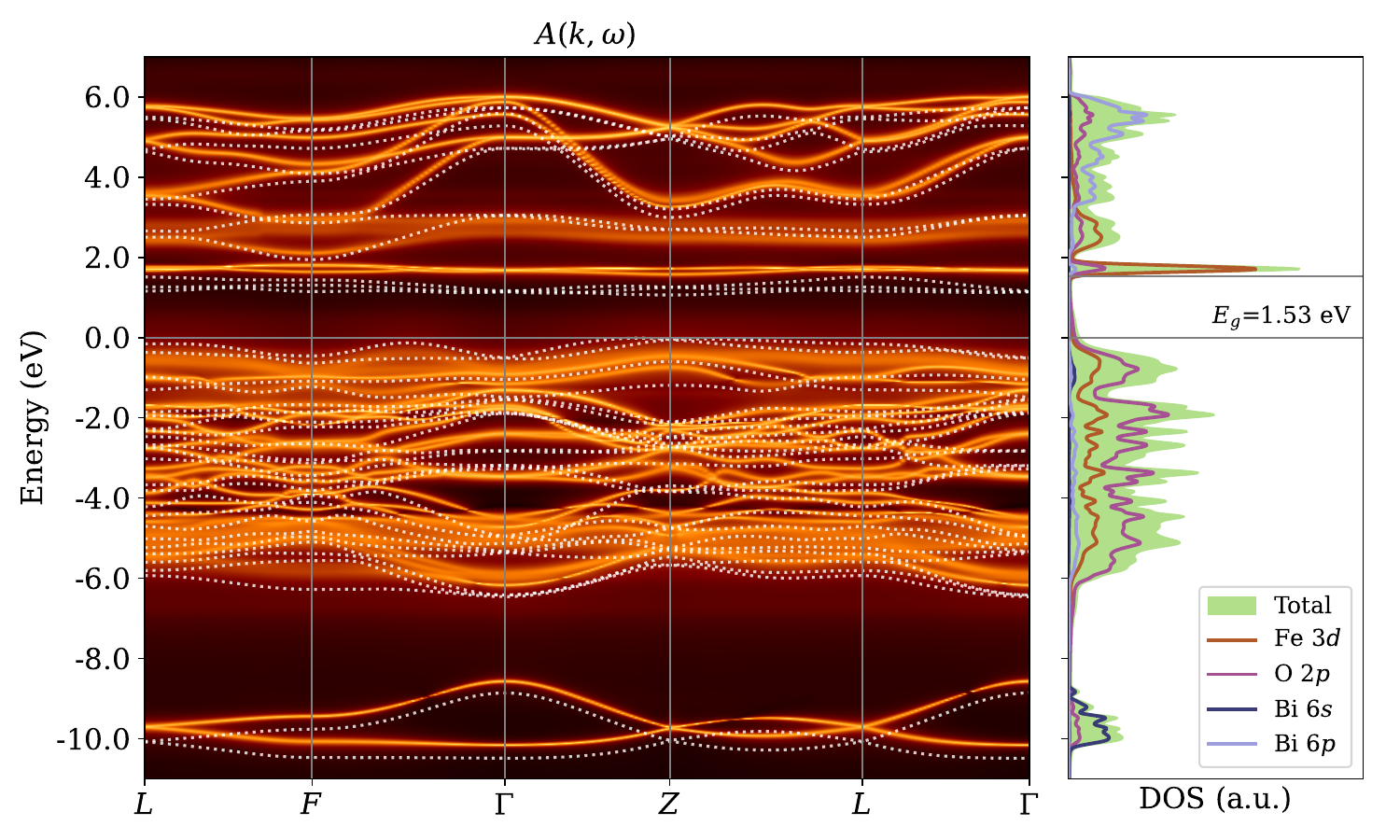}%
        \label{fig:subfigB}%
    }
    \caption{\justifying (a) Conventional (top) and primitive (bottom) $R3c$ \ce{BiFeO3} unit cells. Bi atoms are purple, and FeO$_6$ octahedra are colored by Fe spin orientation (spin-up: red; spin-down: blue). (b) Band structure from PBEsol (white dashed lines) compared with the DFT+$U(\omega)$ spectral function and projected density of states. Dynamical correlations increase the fundamental gap from \SI{1.03}{\eV} to \SI{1.53}{\eV} and introduce significant incoherent spectral weight.}
    \label{fig:widefigure}
\end{figure*}

More advanced many-body techniques, such as dynamical mean-field theory (DMFT) combined with DFT (DFT+DMFT), offer a more rigorous treatment of local correlations introducing a dynamical self-energy~\cite{kotliar_electronic_2006, georges_dynamical_1996}. DMFT applications to \ce{BiFeO3} and related multiferroics have indeed yielded improved density of states and theoretical spectra, showing better agreement with experiments~\cite{craco_electronic_2019, paul_investigation_2018, shorikov_pressure-driven_2015}. However, fully self-consistent DMFT calculations are computationally very demanding, especially for systems with large unit cells or complex magnetic structures like the spin cycloid in \ce{BiFeO3}.

Recently, a new class of computationally efficient dynamical embedding methods has emerged, bridging the gap between static corrections and full DMFT. These approaches, including the dynamical Hubbard functional ($\text{dynH}$)~\cite{chiarotti_energies_2024,caserta2025dynamicalhubbardapproachcorrelated,chiarotti2025self} and the related DFT+$U(\omega)$ method~\cite{vanzini_towards_2023}, are derived from a localized $GW$ approximation on a correlated subspace. They replace the static Hubbard $U$ with a frequency-dependent interaction $U(\omega)$, typically computed using the random-phase approximation (RPA). The DFT+dynH augments the DFT functional by the dynamical Hubbard  correction, fully generalizing the  rotationally invariant DFT+$U$ functional~\cite{dudarev_electron-energy-loss_1998}. DFT+$U(\omega)$ employs a simplification of this method, correcting only the self-energy, while retaining a computational cost only modestly higher than static DFT+$U$, making it an ideal tool for exploring complex correlated materials.

In this Letter, we apply this dynamical DFT+$U(\omega)$ scheme to antiferromagnetic \ce{BiFeO3} to resolve the failures of static corrections. We demonstrate its effectiveness through a direct comparison of simulated hard X-ray photoelectron spectroscopy (HAXPES) spectra against static DFT+$U$+$V$, and dynamical mean-field theory (DMFT) results, clarifying how different correlation treatments affect the band gap and key spectral signatures.

First, we perform DFT+$U$+$V$ calculations using the PBEsol functional~\cite{perdew_restoring_2008} to reproduce results consistent with prior static treatments~\cite{inizan_study_2018,mazumdar_valence_2016,paul_investigation_2018}. Here, the Hubbard $U$ and intersite $V$ parameters are the ones determined self-consistently via density functional perturbation theory by Inizan~\cite{inizan_study_2018}: $U_{\text{Fe}} = \SI{5.5}{eV}$ (on-site Fe $3d$), $V_{\text{Fe-O1}} = \SI{0.9}{eV}$, and $V_{\text{Fe-O2}} = \SI{0.7}{eV}$ for Fe-O interactions. The calculations yield an optimized G-type antiferromagnetic structure in good agreement with experimental data~\cite{kubel_structure_1990} and prior theoretical work~\cite{inizan_study_2018}. However, analysis of the partial density of states (PDOS) confirms the presence of the spurious Fe $3d$-derived peak around \SI{-7}{eV} (Fig.~\ref{fig:haxpes_expanded} and \textcolor{Plum}{S2} in the Supplemental Material), the known artifact of static Hubbard corrections in this system~\cite{paul_investigation_2018, inizan_study_2018, mazumdar_valence_2016}. 

Our dynamical treatment starts with the calculation of the frequency-dependent screened Coulomb interaction, $W(\omega; \mathbf{r}_1, \mathbf{r}_2)$, evaluated within the collinear spin-polarized RPA on the DFT (PBEsol) wavefunctions. The local matrix elements are obtained with the projection onto a basis of maximally localized Wannier functions (MLWFs) \(|\phi_{I m}^{\sigma}\rangle\)~\cite{marzari_maximally_2012} that define the correlated subspace $\mathcal{C}$ (i.e., Fe $3d$). The MLWFs were constructed targeting the Fe $3d$, O $2p$, and Bi $6p$ states, resulting in a 34-orbital model per spin channel. The on-site interaction in the local basis at a given atomic site $I$ for each spin is then:
\begin{equation}
    U_{I; m_1',m_2',m_1,m_2}^{\sigma}(\omega)
    = \langle \phi_{I m_1'}^{\sigma} \phi_{I m_2'}^{\sigma} | W(\omega) | \phi_{I m_1}^{\sigma} \phi_{I m_2}^{\sigma} \rangle.
\end{equation}
Following Refs.~\cite{caserta2025dynamicalhubbardapproachcorrelated,vanzini_towards_2023, amadon_screened_2014}, we defined a scalar $U(\omega)$ by taking the average over the direct density–density interaction matrix within the correlated subspace $\mathcal{C}$ (Fe $3d$, size $N=5$) and each spin channel:
\begin{equation}
    U_I(\omega) \equiv \frac{1}{2N^2}\sum_{m_1,m_2 \in \mathcal{C};\sigma}
    U_{I; m_1,m_2,m_1,m_2}^{\sigma}(\omega).
    \label{eq:onsite_expanded}
\end{equation}
The average of the spin is justified by noting that the spin-up and spin-down interactions differ very little from one another (see Supplemental Material). Having averaged over the spins, we then correct each Fe site with the same interaction, thus from now on the atomic index on the interaction $U_I(\omega)\equiv U(\omega)$ is dropped.

\begin{figure}[h]
    \centering
    \includegraphics[width=1\linewidth]{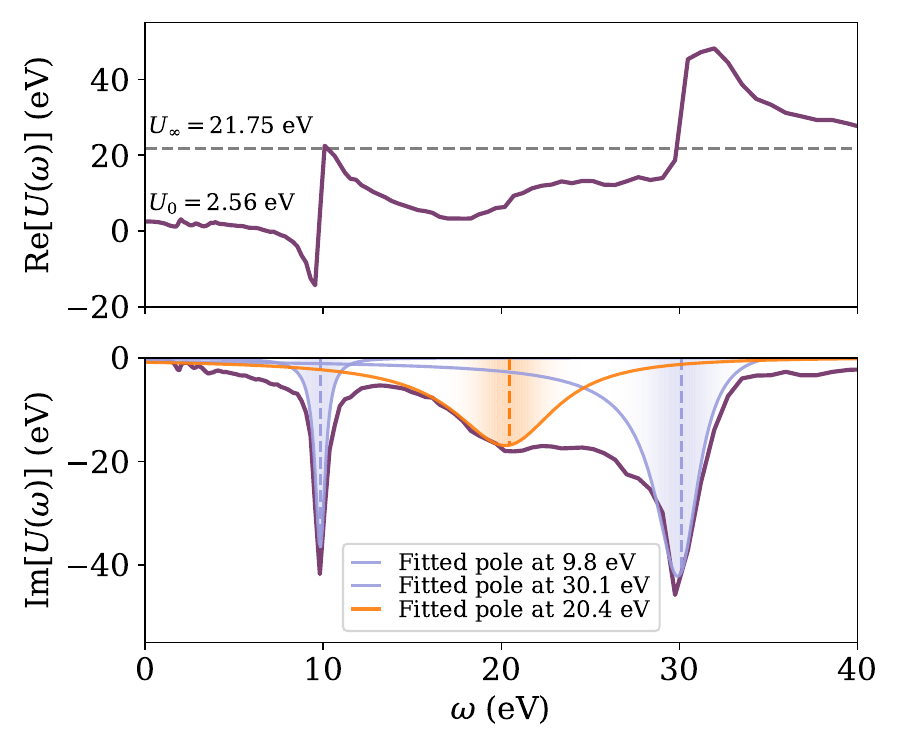}
    \caption{\justifying Real and imaginary parts of the dynamically screened on-site Coulomb interaction $U(\omega)$ for Fe $3d$ orbitals. The static limit is $U_0 = \SI{2.56}{eV}$, and the bare limit is $U_\infty = \SI{21.75}{eV}$.}
    \label{fig:U_expanded}
\end{figure}

The computed frequency-dependent interaction, plotted in Fig.~\ref{fig:U_expanded}, encapsulates the dynamical screening within the Fe $3d$ subspace. In the static limit ($\omega \to 0$), the interaction is strongly screened, yielding an effective $U_0 = \SI{2.56}{eV}$, compared to the bare unscreened value of $U_\infty = \SI{21.75}{eV}$. To obtain a frequency-grid-independent representation and facilitate subsequent DFT+$U(\omega)$ calculations, the RPA result is fitted to a sum-over-poles (SOP) model~\cite{ferretti_greens_2024}. The rich energy dependence is revealed in the imaginary part, $\operatorname{Im}[U(\omega)]$, which represents the spectrum of screening processes. In addition to a continuum of low-energy electron-hole excitations, the spectrum is dominated by three prominent plasmon poles, identified by the SOP fit at \SI{9.84}{eV}, \SI{20.44}{eV}, and \SI{30.11}{eV} within 0--\SI{40}{eV}. This complex structure, which cannot be captured by any single static $U$ value, highlights the crucial role of dynamical correlations. 

As said previously, to incorporate these dynamical effects into the electronic structure, we use an approximate on-site self-energy, $\Delta\Sigma^I_m(\omega)$\cite{vanzini_towards_2023}. This correction assumes correlations are localized on the Fe $3d$ orbitals and that only direct density-density interactions are relevant:
\begin{multline}
    \Delta \Sigma^I_m (\omega) = (\frac{1}{2} - n^I_m)U_\infty +\\
    \left[
        \frac{1}{2} \left(1 + \frac{\omega - \varepsilon^I_m}{\omega_0} \right) - n^I_m
        \right]
    \left(U(\omega - \varepsilon^I_m) - U_\infty\right).
    \label{eq:self-energy_expanded}
\end{multline}
Here, $n^I_m$ is the initial occupation of the localized orbital $m$ on the Fe site $I$, and the parameter $\omega_0$ represents a characteristic energy for the screening, obtained by approximating the full spectrum of neutral excitations with a single effective frequency~\cite{vanzini_towards_2023,berger_ab_2010}. We set $\omega_0 = \SI{20}{eV}$, a choice justified by the analysis of $\operatorname{Im}[U(\omega)]$, where this value corresponds to the broadest plasmon pole (at \SI{20.44}{eV}) and is also approximately the average of the three main poles. $\varepsilon^I_m$ are the Kohn-Sham eigenvalues in the local basis $m$, analogous to the local occupations $n^I_m$, used in the DFT+U context, and they stem from the frequency dependent extension of the formalism~\cite{vanzini_towards_2023}. This self-energy can be decomposed of exchange and correlation parts, plus a double-counting correction term in the fully localized limit (FLL), with a static value of $U_\infty = \SI{21.75}{eV}$. See Chiarotti et al. for details~\cite{chiarotti_energies_2024}. The final electronic structure is obtained by computing the spectral function as, $A_{n\mathbf{k}}(\omega) = -\frac{1}{\pi} \text{Im} [(\omega - \varepsilon_{n\mathbf{k}} - \sum_{I,m} |\langle I,m|n\mathbf{k}\rangle|^2 \Delta \Sigma^I_m(\omega))^{-1}]$, i.e.,  by projecting back the local self-energy onto the Kohn-Sham space and correcting Kohn-Sham  eigenvalues ~\cite{vanzini_towards_2023}.

Our DFT+$U(\omega)$ calculations yield two crucial improvements over static methods. First, the fundamental band gap is corrected to \SI{1.53}{eV} — a value that rectifies both the underestimation by plain DFT (\SI{1.03}{eV}) and the overestimation by DFT+$U$+$V$ (\SI{2.61}{eV}). This result aligns well with experimental reports, which suggest an indirect gap of approximately \SI{1.2}{eV} to \SI{1.9}{eV}~\cite{schmidt_anisotropic_2015,fruth_preparation_2007,mcdonnell_photo-active_2013,ayala_study_2022,gujar_nanocrystalline_2007}. Second, and most critically, our dynamical approach resolves the primary failure of static mean-field corrections. The spectral function (Fig.~\ref{fig:subfigB}) shows that the spurious, localized Fe $3d$ peak near \SI{-7}{eV} is entirely eliminated. Instead, the Fe $3d$ spectral weight is now correctly distributed across the upper valence band, from the Fermi level down to around \SI{-6}{eV}, reflecting its strong hybridization with O $2p$ and Bi $6p$ states.

\begin{figure}[h]
    \centering
    \includegraphics[width=1\linewidth]{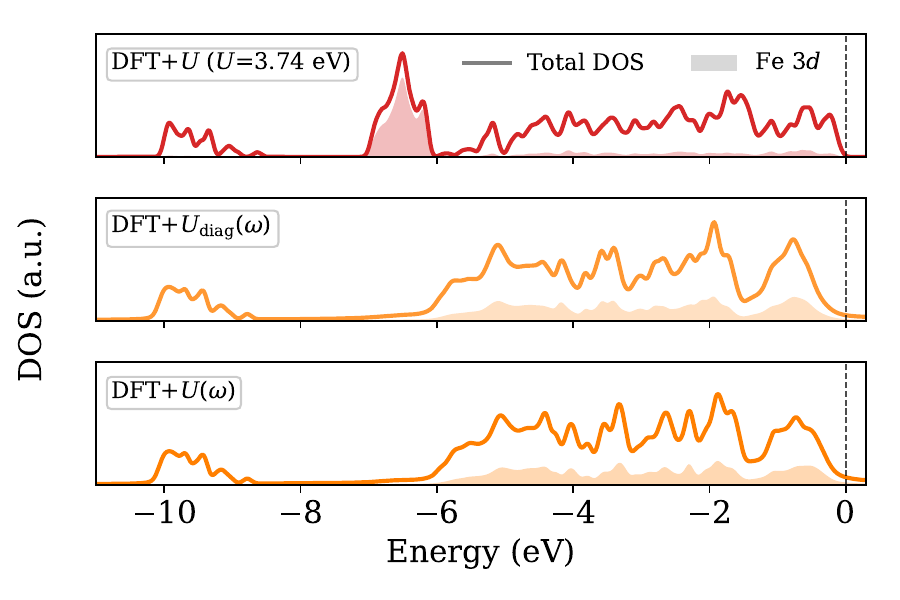}
    \caption{\justifying Disentangling static vs. dynamical effects on the Fe $3d$ PDOS. A static DFT+$U$ calculation using $U=\SI{3.74}{eV}$ (top) produces a spurious \ce{Fe} peak. In contrast, a dynamical $\text{DFT}+U_{\text{diag}}(\omega)$ calculation averaging only the diagonal elements (middle), which has the same large static limit but includes dynamics, correctly eliminates this artifact, closely matching the result from DFT+$U(\omega)$ with full averaging (bottom).}
    \label{fig:dos_diag}
\end{figure}

One might argue that this correction is merely an artifact of the small static limit $U_0 = \SI{2.56}{eV}$, which is below the threshold needed to induce the spurious peak in a static calculation (see Supplemental Material). To unambiguously disentangle the roles of the static limit versus the frequency dependence, we performed a decisive numerical experiment. We construct an alternative interaction, $U_{\text{diag}}(\omega)$, simply obtained by a different average over the local manifold. For this, we use only the orbital diagonal elements (see Supplemental Material). This preserves a nearly identical dynamical profile but yields a considerably larger static limit of $U_{\text{diag}}(0) = \SI{3.74}{eV}$. As shown in Fig.~\ref{fig:dos_diag}, a static DFT+$U$ calculation with this larger value produces a pronounced artifact peak. The dynamical $\text{DFT}+U_{\text{diag}}(\omega)$ calculation, however, completely suppresses it. This result provides compelling evidence that the redistribution of the spurious peak is a genuine consequence of the dynamical screening and not simply due to a small effective Hubbard $U$.

Following other theoretical work~\cite{paul_investigation_2018}, to provide a more direct comparison with experiment, we simulated HAXPES spectra from the calculated valence spectral function. The simulation accounts for atomic orbital photoionization cross-sections at an incident energy of \SI{2100}{\eV}, weighting the orbital-projected DOS with tabulated values~\cite{mazumdar_valence_2016} and applying a Gaussian broadening to match instrumental resolution. Figure~\ref{fig:haxpes_expanded} compares our results from DFT+$U(\omega)$ and DFT+$U$+$V$ against experimental data~\cite{mazumdar_valence_2016, paul_investigation_2018} and DFT+DMFT~\cite{paul_investigation_2018}.

\begin{figure}[h]
    \centering
    \includegraphics[width=1\linewidth]{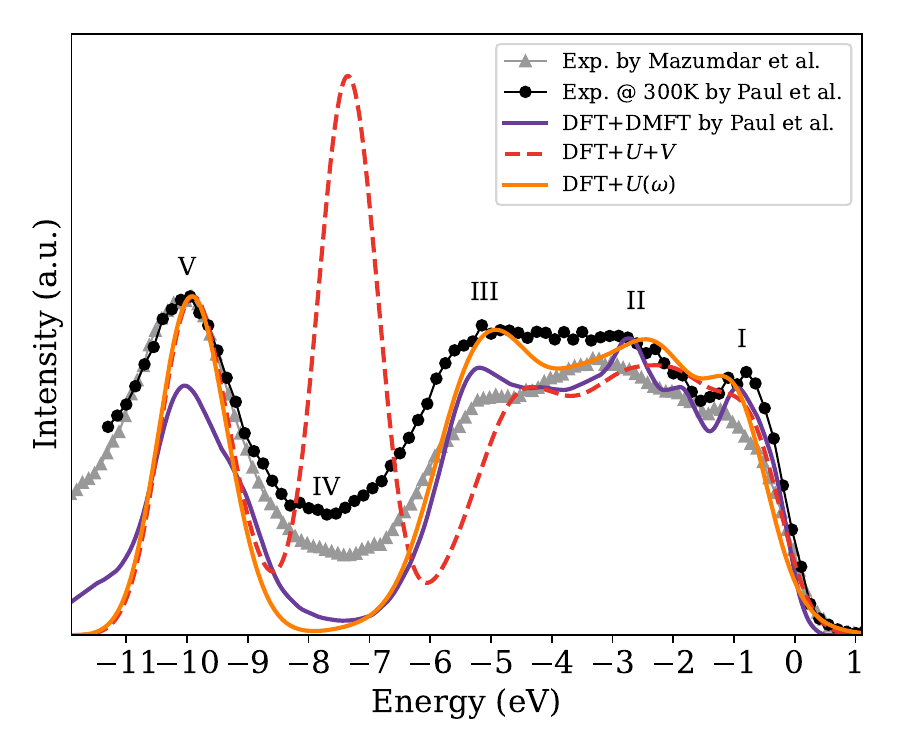}
    \caption{\justifying Comparison of simulated HAXPES spectra with experimental data from Refs.~\cite{mazumdar_valence_2016,paul_investigation_2018} (points). The DFT+$U(\omega)$ spectrum (orange) provides the best overall agreement with experiment, eliminating the spurious peak of static DFT+$U$+$V$ (red, dashed) and matching or exceeding the accuracy of DFT+DMFT (purple).}
    \label{fig:haxpes_expanded}
\end{figure}

\begin{figure}[h]
    \centering
    \includegraphics[width=1\linewidth]{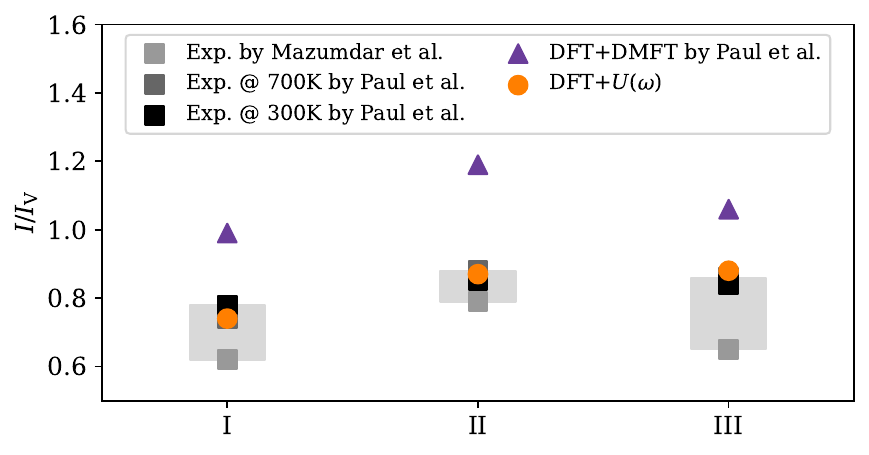}
    \caption{\justifying Relative intensity of HAXPES features. The plot shows the intensity ratios of valence features I, II, and III relative to peak V. The DFT+$U(\omega)$ results fall within the experimental range (grey bars), while the DFT+DMFT~\cite{paul_investigation_2018} method consistently overestimates these ratios.}
    \label{fig:haxpes_intensity}
\end{figure}

The DFT+$U(\omega)$ spectrum demonstrates remarkable agreement with the experimental data. In contrast to the static DFT+$U$+$V$ result, which incorrectly places a pronounced peak at the position of feature IV, our dynamical calculation correctly shows a valley in this region and entirely eliminates the spurious Fe $3d$ artifact. The energy positions of all key spectral features — peaks I, II, III, and the deeper-lying peak V — are accurately reproduced, matching both experimental measurements and DFT+DMFT calculations. Peak positions and intensities were determined by fitting the spectra with Voigt functions, as detailed in the Supplemental Material. Notably, our method surpasses the DMFT result in one key aspect: it more faithfully captures the relative intensity of the Bi $6s$-related peak V with respect to the main valence band features (I–III) (Fig.~\ref{fig:haxpes_intensity}). This success in reproducing the experimental lineshape highlights the predictive power of the DFT+$U(\omega)$ approach. We confirm all these findings by using the (more complete) DFT+dynH approach and discuss the comparison in the Supplemental Material.

In conclusion, we have shown that incorporating dynamical screening via a frequency-dependent Hubbard interaction resolves the long-standing failure of static mean-field methods to describe the electronic structure of \ce{BiFeO3}. Our DFT+$U(\omega)$ calculations eliminate the spurious deep-valence Fe $3d$ peak, correct the fundamental band gap, and reproduce experimental HAXPES spectra with a fidelity that matches or exceeds that of far more computationally demanding DMFT calculations. This work not only clarifies the crucial role of dynamical correlations in multiferroics but also establishes DFT+$U(\omega)$ as a predictive, first-principles tool with a compelling balance of accuracy and efficiency. Its success here paves the way for reliable, large-scale simulations of a wide range of complex correlated materials where such effects are paramount.

\bibliographystyle{apsrev4-2}
\bibliography{Bibliography}

\onecolumngrid

\ifarXiv
    

\section*{Supplementary material}

**DFT + U + V**

### Computational Details

Calculations are performed using `Quantum ESPRESSO` [S1] with parameters from Ref. [S2]: $U = 5.45$ eV, $V_1 = 0.94$ eV (Fe-O1), and $V_2 = 0.72$ eV (Fe-O2), determined via self-consistent Density Functional Perturbation Theory (DFPT) [S2, S3]. We employ the PBEsol functional [S4] and SSSP pseudopotentials [S5] with wavefunction and charge density cutoffs of 90 Ry and 1080 Ry, respectively. A cold smearing of 0.02 Ry [S6] is used for SCF calculations on a $9 \times 9 \times 9$ Monkhorst-Pack grid [S7], while NSCF calculations use the tetrahedron method on a $14 \times 14 \times 14$ grid [S8].

### Electronic Structure

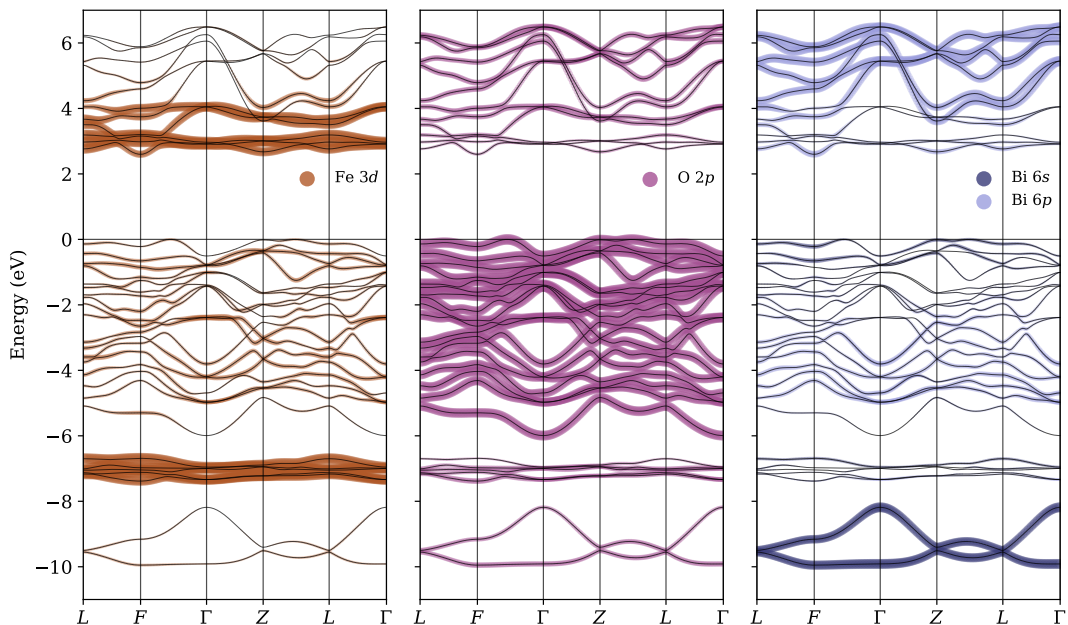

FIG. S1: Projected band structure calculated using the DFT+$U$+$V$ method with $U = 5.45$ eV, $V_1 = 0.94$ eV, and $V_2 = 0.72$ eV. The left, middle, and right panels show the orbital projections onto Fe, O, and Bi atoms, respectively. The valence band maximum is set to zero energy, and the resulting indirect band gap is 2.61 eV.

The static DFT + $U$ + $V$ calculation overestimates the band gap relative to most experimental measurements. Specifically, an indirect gap of 2.61 eV is obtained. Although this value appears reasonable *a priori* for an oxide with correlation effects, it slightly exceeds the commonly cited experimental range of ~1.2 eV to 1.9 eV for BiFeO$_3$ [S9–S13].

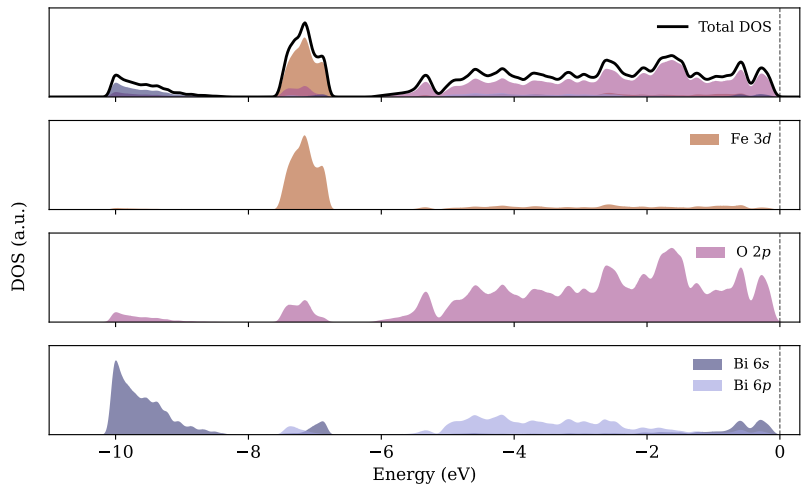

FIG. S2: Total and projected density of states (PDOS) of $BiFeO_3$ from DFT + $U$ + $V$. The PDOS highlights the orbital character of the valence band, showing the spurious, isolated Fe $3d$ peak near $-7.6\,\mathrm{eV}$.

A crucial point of concern can be seen in the partial Fe $3d$ density of states (PDOS). As indicated in Fig. S2, a pronounced unphysical peak appears around $-7\,\mathrm{eV}$, mirroring the issue reported by Paul *et al.* using the simpler LSDA+$U$ scheme [S14]. Its persistence in our DFT+$U$+$V$ calculations highlights that static Hubbard corrections, even with additional intersite terms, are not sufficient to fully capture the intricate many-body effects in $BiFeO_3$.

## DFT+$U(\omega)$

### Computational Details

DFT+$U(\omega)$ calculations employ PBEsol [S4] and ONCVPSP pseudopotentials (PseudoDojo v0.4.1) [S15]. We use energy cutoffs of $106\,\mathrm{Ry}$ (wavefunction) and $424\,\mathrm{Ry}$ (density), with $0.007\,\mathrm{Ry}$ cold smearing [S6]. The SCF step uses a $10 \times 10 \times 10$ grid [S7], while NSCF steps use a $6 \times 6 \times 6$ mesh.

We employ a $dp$-model (Fe $3d$, O $2p$, Bi $6p$) with 34 maximally localized Wannier functions (MLWFs) [S16]. $U(\omega)$ is computed with a $10\,\mathrm{Ry}$ dielectric cutoff, 100 bands, and 200 frequency points distributed logarithmically [S17].

The workflow proceeds as follows: `Quantum ESPRESSO` [S1] generates Bloch wavefunctions, which a customised version of `RESPACK` [S17] uses to construct MLWFs (`calc_wannier`) and compute the frequency-dependent dielectric function via spin-polarized RPA (`calc_chiqw`). The on-site $U(\omega)$ is obtained by projecting the screened interaction onto the Wannier basis (`calc_w3d`) and averaging the matrix elements. Finally, $U(\omega)$ is fitted to a sum-over-poles form [S18] for the self-energy correction [S19], and spectral functions are computed via upfolding.

### Averaged $U(\omega)$

The DFT+$U(\omega)$ self-energy is obtained by approximating the convolution between the local Green's function and the local screened Coulomb interaction of the system. For the complete derivation, we refer to see [S19]. As the screened Coulomb interaction is averaged over a Maximally Localized Wannier Function basis, the spin-dependency of the matrix elements

$$U_{I;m_1',m_2',m_1,m_2}^{\sigma}(\omega) = \langle \phi_{Im_1'}^{\sigma} \phi_{Im_2'}^{\sigma} | W(\omega) | \phi_{Im_1}^{\sigma} \phi_{Im_2}^{\sigma} \rangle. \tag{S1}$$

stem from the spin dependency of the MLWFs $\phi_{Im}^{\sigma}$, which in turns stems from the spin dependency of the Kohn-Sham states $\psi_{n\mathbf{k}}^{\sigma}$. Thus the spin-dependency of the interaction might be non-trivial, if the spin-down and spin-up Kohn-Sham states comprising the local manifold significantly differ from each other in charge distribution around the Iron ion. In Fig. S4 we show the spin-up and spin-down component of the screened interaction projected onto one of the two Fe atoms of the lattice (by antiferromagnetic symmetry, the other Fe atom has identical but swapped up and down MLWFs). As it can be seen, the up and down interactions have identical pole structures (Imaginary part)

and differ very little (less than 0.2 eV) in the real part across the low-energy range. This justifies the spin-average of the local matrix elements S1, leading thus to the total average for each Iron site used in this work. Figure S3 shows $d_{x^2-y^2}$-like Wannier functions in real space, centered at atomic site Fe1 and Fe2, for spin up and down channels, where the similarity of the up and down orbitals per site can be appreciated.

$$U_I(\omega) \equiv \frac{1}{2N^2} \sum_{m_1,m_2 \in \mathcal{C};\sigma} U^{\sigma}_{I;m_1,m_2,m_1,m_2}(\omega).$$
(S2)

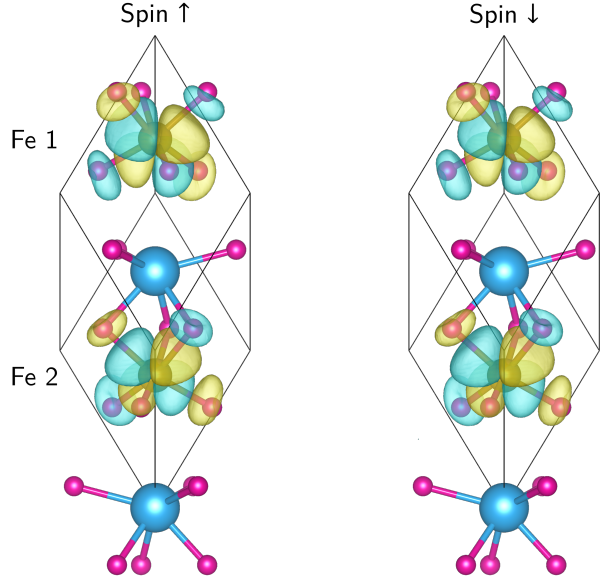

FIG. S3: Real-space $d_{x^2-y^2}$-like Wannier functions centered at atomic site Fe1 and Fe2 for spin up and down channels. The spin-up WF on Fe1 is identical in shape to the spin-down WF on Fe2, ensured by spatial and time-reversal symmetry. The gauge-invariant spread of the spin-up WF differs from the corresponding spin-down WF by only about 12%, confirming that the local orbital shapes remain nearly identical across different spin sector at the same site. This can be understood as the spin-dependent part of the exchange correlation is small and perturbative.

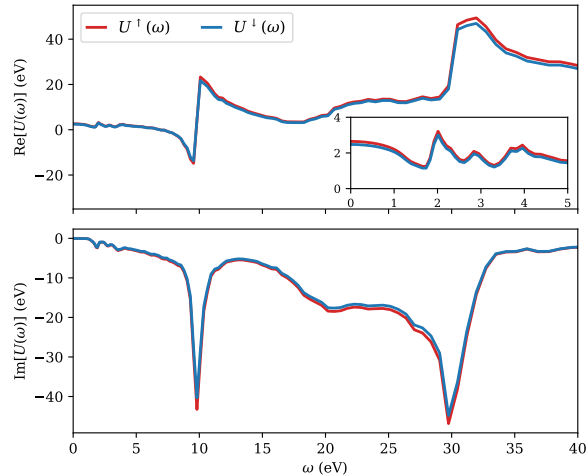

FIG. S4: Comparison of the frequency-dependent local Coulomb interaction in spin-up (red) and spin-down channels (blue). They are almost identical, justifying to have an averaged spin-independent $U(\omega)$.

## Dynamical Correlations $U(\omega)$ in BiFeO$_3$

Since $U(\omega)$ will be used in later DFT+$U(\omega)$ or DFT+dynH calculations, it is beneficial to make $U(\omega)$ inpedependent on the RPA energy grid by fitting it on a sum-over-poles (SOP) representation [S18] . Fig. S5 compares the $U(\omega)$ from RPA (black dashed lines) to its SOP fit (orange lines). The SOP procedure attempts to capture the most relevant peaks in both $\mathrm{Re}\,U(\omega)$ and $\mathrm{Im}\,U(\omega)$ with a small set of damped oscillators. The agreement is close across the entire $0-40\,\mathrm{eV}$ range, which are necessary to describe the low energy frequency dependence of the onsite interaction in BiFeO$_3$.

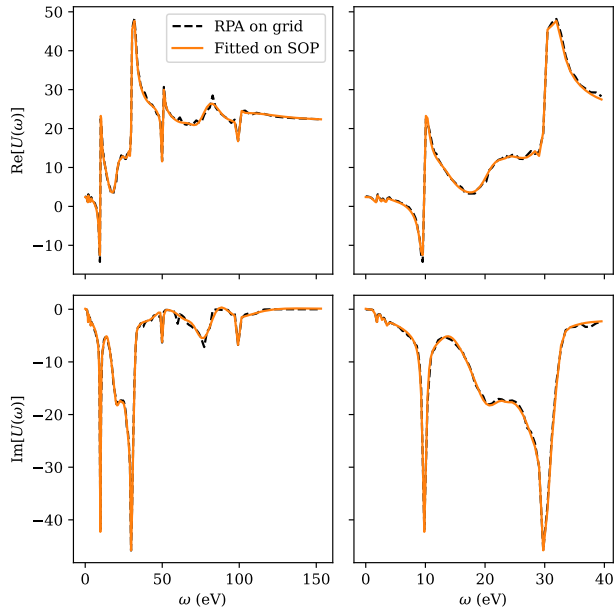

FIG. S5: Real and imaginary parts of the on-site screened interaction $U(\omega)$. The raw RPA calculation (black dashed line) is fitted to a sum-over-poles (SOP) representation (orange line), which is used in later DFT+$U(\omega)$ and DFT+dynH calculations.

## DISTINGUISHING DYNAMICAL FROM STATIC CORRELATION EFFECTS

A central failure of static Hubbard corrections in BiFeO$_3$ is the appearance of a spurious, localized Fe $3d$ peak in the valence band around $-7\,\mathrm{eV}$, an artifact that is however absent in the plain DFT calculation (Fig. 1 in the main text). To understand the origin of this feature, we perform a series of static DFT+$U$ calculations, varying the Hubbard $U$ from 0 to the self-consistent DFPT value of $5.01\,\mathrm{eV}$. As shown in Fig. S6, the unphysical peak emerges for $U > 2\,\mathrm{eV}$ and becomes progressively more pronounced, separating from the main valence band as $U$ increases. This behavior is expected from the Hubbard model, which enforces localization with increasing $U$, but it contradicts experimental HAXPES data that suggest dynamical hybridization between Fe $3d$ and O $2p$ orbitals.

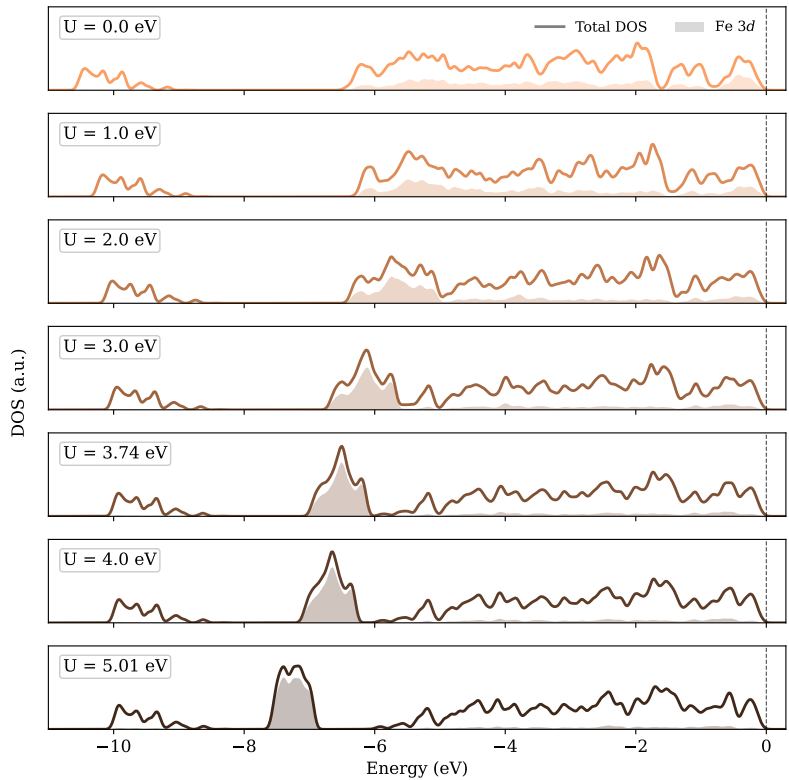

FIG. S6: Total DOS and PDOS projected onto the Fe $3d$ orbitals from static DFT + $U$ calculations with increasing $U$. The spurious Fe peak emerges for $U$ greater than $2\,\mathrm{eV}$. $3.74\,\mathrm{eV}$ is the static limits for $U_{\mathrm{diag}}(0)$ and $5.01\,\mathrm{eV}$ is the Hubbard parameter determined from self-consistent DFPT calculations.

This raises a critical question: does the DFT+$U(\omega)$ correction arise from true dynamical screening or merely its small static limit ($U_0 = 2.56\,\mathrm{eV}$)? To test this, we constructed a hypothetical $U_{\mathrm{diag}}(\omega)$ by averaging only the diagonal elements of the local interaction matrix:

$$U_{I;\mathrm{diag}}(\omega) \equiv \frac{1}{2N} \sum_{m \in \mathcal{C}; \sigma} U^{\sigma}_{I;m,m,m,m}(\omega). \tag{S3}$$

As in the main text, we omit the atomic index here and demote $U_{\mathrm{diag}}(\omega) \equiv U_{I;\mathrm{diag}}(\omega)$. This yields a nearly identical dynamical profile but a significantly larger static limit, $U_{\mathrm{diag}}(0) = 3.74\,\mathrm{eV}$ (see Fig. S7), well within the range where static DFT+$U$ produces the spurious peak.

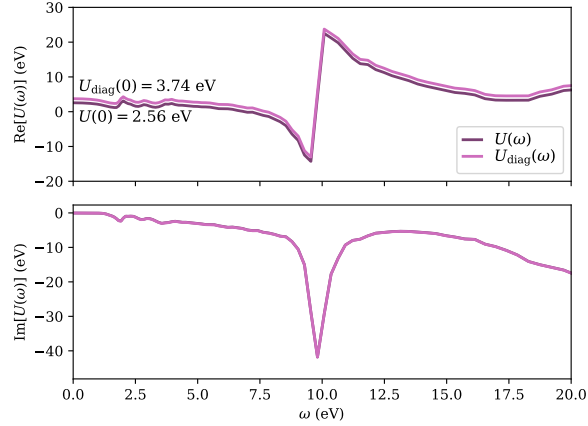

FIG. S7: Comparison of the frequency-dependent local Coulomb interaction from element-wise averaging ($U(\omega)$, dark plum) and diagonal averaging ($U_{\mathrm{diag}}(\omega)$, magenta). While the dynamical profiles are similar, the static limits are $U(0) = 2.56\,\mathrm{eV}$ and $U_{\mathrm{diag}}(0) = 3.74\,\mathrm{eV}$.

The test is decisive. Static DFT+$U$ ($U = 3.74\,\mathrm{eV}$) produces the spurious Fe peak (Fig. S6) and a mismatched HAXPES spectrum (Fig. S8). In contrast, dynamical DFT+$U_{\mathrm{diag}}(\omega)$ — despite the same large static limit — yields results (Fig. S9) in excellent agreement with the full DFT+$U(\omega)$ calculation and experiments. This confirms that the correction is driven by dynamical screening, not a reduced static $U_0$.

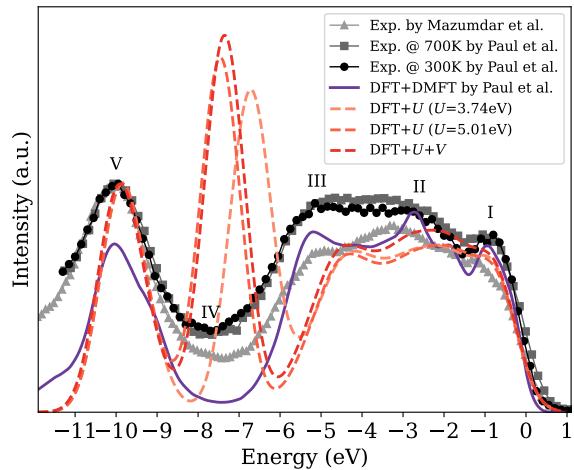

FIG. S8: Simulated HAXPES spectra from static calculations. All static treatments fail to reproduce experimental features, showing a pronounced Fe peak where a valley (feature IV) should be.

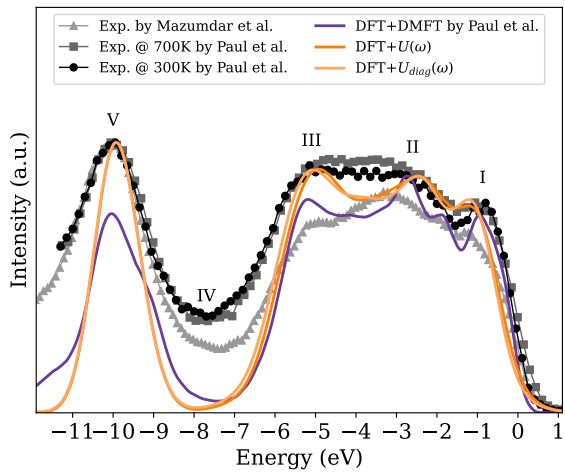

FIG. S9: Simulated HAXPES from dynamical DFT$+U(\omega)$ and DFT$+U_{\mathrm{diag}}(\omega)$ calculations. Both methods eliminate the spurious static peak and show excellent agreement with experiments.

## VOIGT-CURVATURE ANALYSIS OF SPECTRAL FEATURES

Quantitative comparison of HAXPES spectra is challenging because features I, II, and III appear as broad shoulder-like structures rather than sharp peaks. To address this, we employ a two-step analysis. First, we fit each spectrum with a sum of Voigt functions to obtain a smooth, twice-differentiable analytic representation $f(E)$ that filters high-frequency noise.

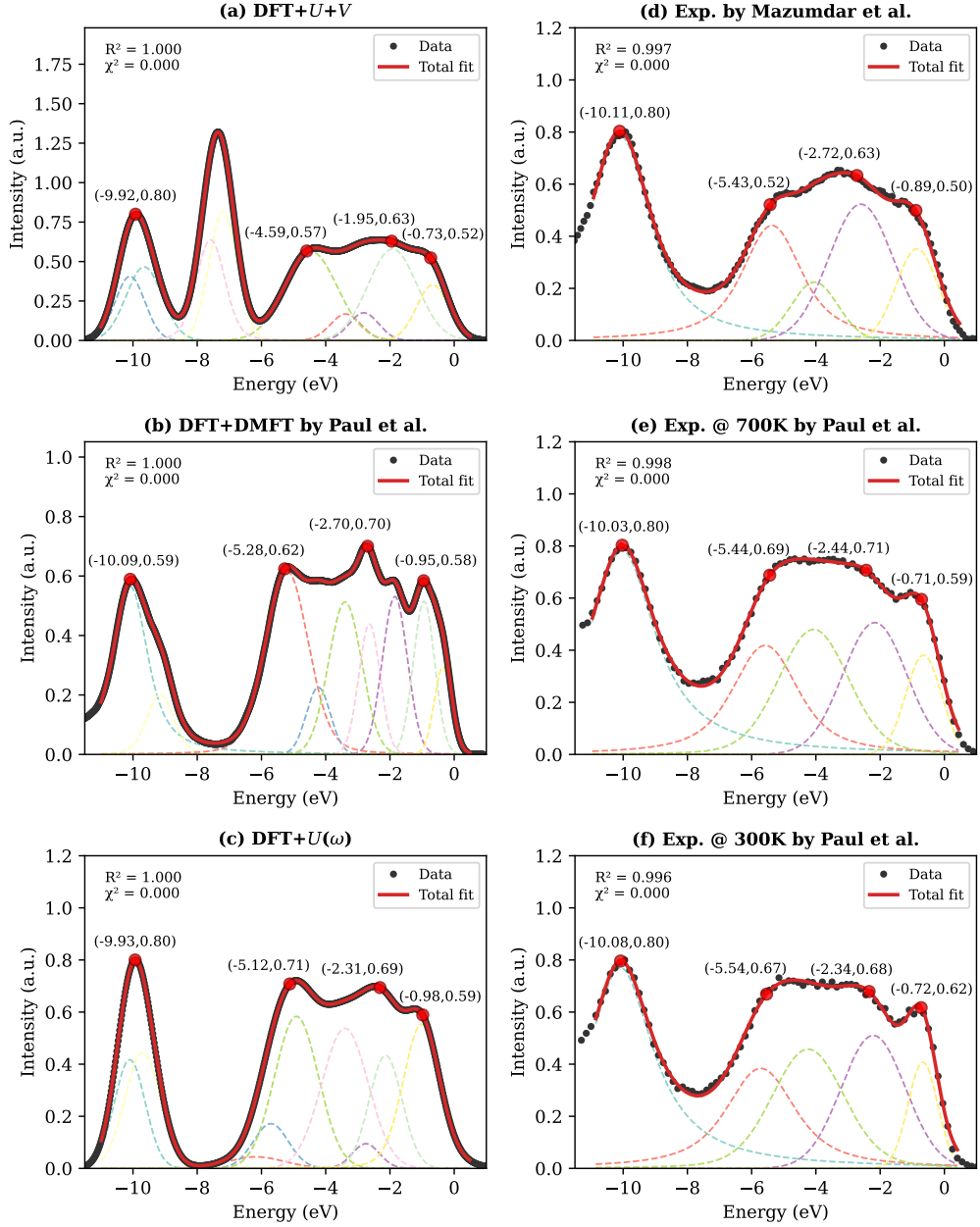

FIG. S10: Voigt function fits to HAXPES spectra. Raw data (points) are overlaid with the smooth Voigt function fit (red line) for experimental and theoretical spectra. The features identified via curvature analysis are highlighted with red points (I–V).

Second, we identify spectral features by locating points of maximum curvature, $\kappa(E) = |f''(E)|/[1 + (f'(E))^2]^{3/2}$. Geometrically, these correspond to points where the osculating circle has the smallest radius. This method provides a robust, consistent metric to determine the position and intensity of broad spectral features across both experimental and theoretical datasets.

# FAILURE OF HIGHER STATIC FUNCTIONALS

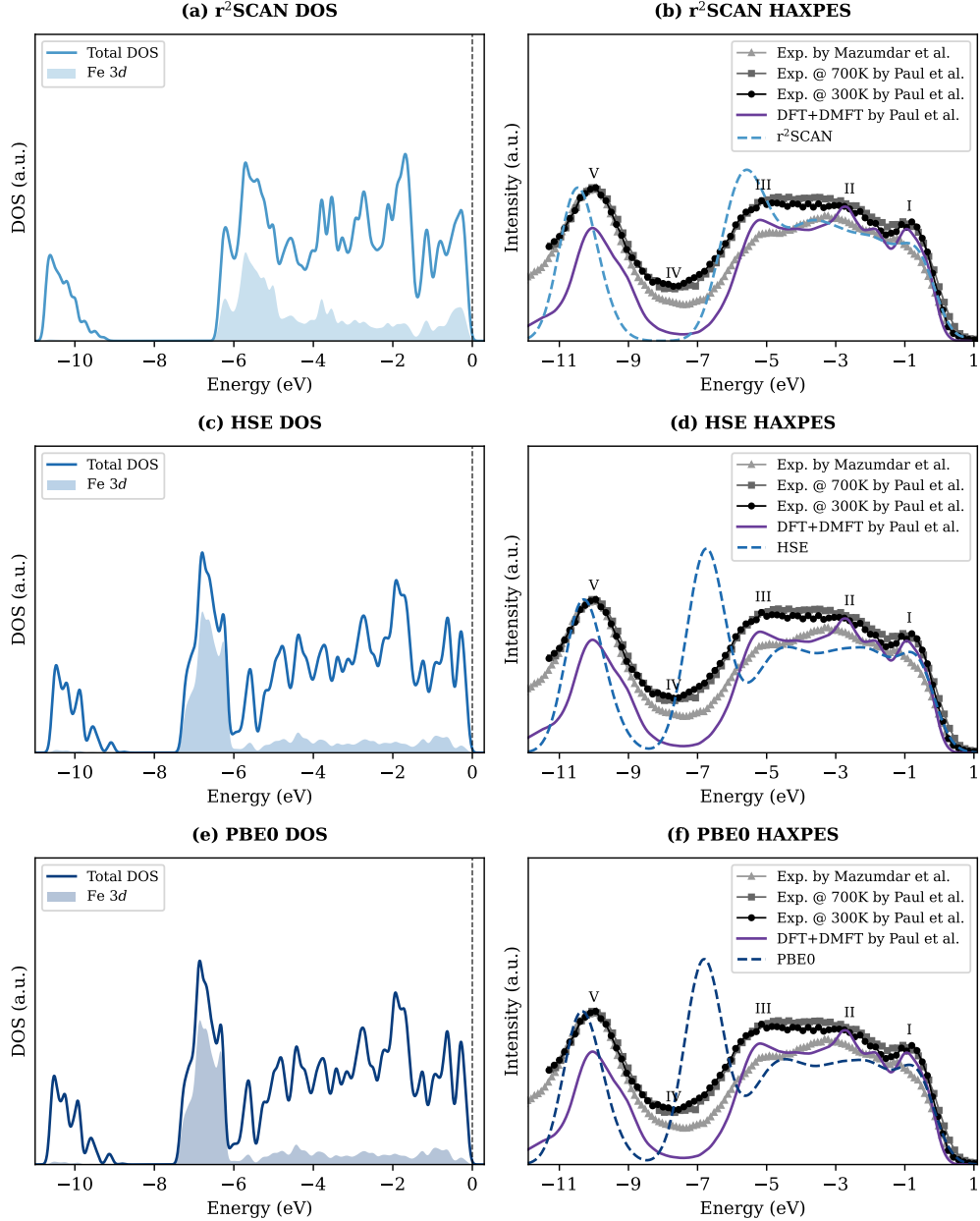

FIG. S11: Failure of advanced static exchange-correlation functionals (r²SCAN, HSE, and PBE0). All three functionals fail to reproduce the experimental spectrum, producing artifacts similar to those seen in static DFT+$U$.

We also investigate whether advanced static functionals can resolve the electronic structure deficiencies. We test the meta-GGA r²SCAN and hybrid functionals HSE and PBE0 [S20–S22]. As shown in Fig. S11, all fail to correct the discrepancies with experiment.

r²SCAN mimics a DFT+$U$ calculation with a small effective $U \approx 2\,\text{eV}$, incorrectly predicting feature III to be more intense than feature V. The hybrid functionals (HSE, PBE0) strongly over-localize the Fe $3d$ states, creating a sharp spurious peak deep in the valence band, analogous to DFT+$U$ with $U \approx 4\,\text{eV}$. These results confirm that static treatments — regardless of the functional's complexity — cannot capture the necessary dynamical screening effects in BiFeO₃.

## COMPARISON WITH DYNAMICAL HUBBARD FUNCTIONAL (DYNH) APPROACH

To validate our findings, we perform calculations using the dynamical Hubbard functional (dynH) approach [S23, S24]. Unlike the one-shot self-energy correction used in our main work, DFT+dynH solves the Dyson equation via an algorithmic-inversion method (AIM-SOP) [S25], offering a rigorous, non-perturbative treatment of the self-energy:

$$\Sigma_{\text{dynH}}^{\sigma,I}(\omega) = 2\pi i \frac{\partial \Phi_{\text{dynH}}[G_I^\sigma]}{\partial G_I^\sigma} = -\int d\omega' \, U_I(\omega') \, G_I^\sigma(\omega + \omega') + \frac{U_I^\infty}{2}. \tag{S4}$$

The results (Fig. S12) confirm our central conclusion: DFT+dynH eliminates the spurious Fe peak and accurately reproduces the spectral feature positions. The lower spectral intensity of features I-III in DFT+dynH compared to experiment likely reflects the absence of error cancellation present in the simpler one-shot approach (where approximate self-energy errors may offset PBEsol deficiencies). However, DFT+dynH more accurately captures the lineshape of features I-III, underscoring the physical robustness of the dynamical treatment.

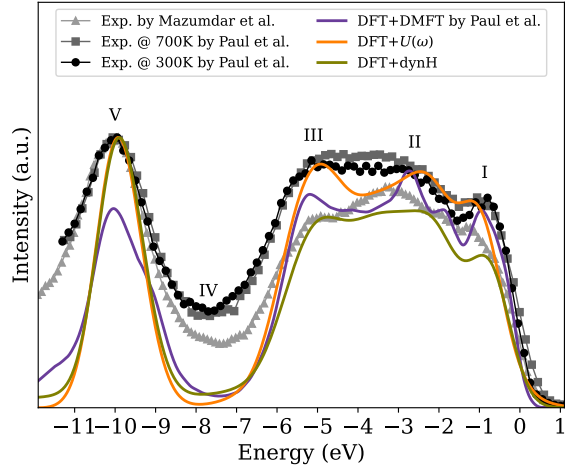

FIG. S12: Comparison of the simulated HAXPES spectrum from the DFT + dynH method (green) with the simpler DFT+$U(\omega)$ result (orange) and experimental data (black and grey scatters). The DFT+dynH approach validates the peak positions but yields lower spectral intensity for features I-III.

---

\* These authors contributed equally to this work.

† nicola.marzari@epfl.ch


\fi

\twocolumngrid

\end{document}